\documentclass[a4paper,twoside]{article}
\newcommand{\affil}[1]{$^{\rm #1}$}
\usepackage[authoryear]{natbib}
\bibpunct{(}{)}{;}{a}{}{,}
\usepackage{graphicx}
\usepackage{hyperref}
\date{} 
\newcommand{\mnu}{$\sum m_\nu$}
\newcommand{\Neff}{$N_\mathrm{eff}$}
\newcommand{\hh}{$H_0$}
\newcommand{\omm}{$\Omega_\mathrm{m,0}$}

\newcommand{\eV}{\, \textrm{\,eV}}
\newcommand{\MeV}{\, \textrm{\,MeV}}

\newcommand{\Mpc}{\textrm{\,Mpc}}

\newcommand{\s}{\mathrm{\,s}}
\newcommand{\km}{\, \mathrm{\,km}}

\newcommand{\LCDM}{$\Lambda$CDM}

\newcommand{\He}{$^4$He}

\newcommand{\jcap}{Journal of Cosmology and Astroparticle Physics}
\newcommand{\prc}{Phys. Rev. C}
\newcommand{\prd}{Phys. Rev. D}
\newcommand{\apjl}{Astrophysical Journal Letters}
\newcommand{\apj}{Astrophysical Journal}
\newcommand{\apjs}{Astrophysical Journal Supplement}
\newcommand{\mnras}{Monthly Notices of the Royal Astronomical Society}

\newcommand\figref[1]{%
Fig.~\ref{fig:#1}}

\newcommand\secref[1]{%
Sec.~\ref{sec:#1}}

\title{\large\bf\flushleft What is half a neutrino? Reviewing cosmological constraints on neutrinos and dark radiation}
\author{\parbox{\textwidth}{\flushleft
\vspace{-0.5cm}
{\it S.~Riemer-S\o{}rensen\affil{A, B}, D.~Parkinson\affil{A}, T.~M.~Davis\affil{A}}\\
\vspace{0.4cm}
{\small \affil{A}\,School of Mathematics and Physics, University of Queensland, St Lucia, Brisbane 4072, Queensland, Australia}\\
{\small \affil{B}\,Email: signe@physics.uq.edu.au}}}
\begin{document}
%
\maketitle
%
%
\small{\bf Abstract: }

Neutrinos are one of the major puzzles in modern physics. Despite measurements of mass differences, the Standard Model of particle physics describes them as exactly massless. Additionally, recent measurements from both particle physics experiments and cosmology indicate the existence of more than the three Standard Model species. Here we review the cosmological evidence and its possible interpretations.
 


\section{Introduction}
Neutrinos are the lightest of the known massive particles and one of the least understood. The cosmic neutrino background has a number density of $112\, \mathrm{cm}^{-3}$ (per specie), which means that billions of neutrinos pass through our bodies every second, so they are by no means rare. Yet we do not know their basic properties, such as their mass and how many species there are.

\subsection{Particle physics neutrinos}
The discovery of the Higgs boson at the Large Hadron Collider (LHC) consolidates the power of the Standard Model of particle physics \citep{ATLAS:2012, CMS:2012}. However, there are still unsolved issues in the neutrino sector, with their masses remaining unexplained and uncertainty about the actual number of species. 

The Standard Model contains three species of mass-less neutrinos that interact only through the weak interaction. By measuring the decay width of the Z$^0$ boson the Large Electron Positron Collider (LEP) measured the number of weakly interacting neutrinos to be $N_\nu = 2.92 \pm 0.05$ \citep{Beringer:2012}, corresponding to the three generations of leptons namely the electron, the muon and the tau particles.

Observations of neutrino disappearance, e.g. from $\nu_e$ sources \citep{Cleveland:1998, Fukuda:1998}, and neutrino appearances e.g. in $\nu_\mu$ beams \citep{Abe:2011}, have lead to the framework of neutrino mixing/oscillations, where the interaction eigenstates ($\nu_e,\nu_\mu,\nu_\tau$) are described as superpositions of mass eigenstates ($\nu_1, \nu_2, \nu_3$). Global fits to data from experiments using solar, atmospheric, and reactor neutrinos have measured mass differences between the three species to be $\Delta m_{32}^2 = |(2.43^{+0.12}_{-0.08}) \times10^{-3}|\eV^2$ and $\Delta m_{21}^2 = (7.50\pm0.20)\times10^{-5}\eV^2$ \citep{Beringer:2012} thus requiring at least two mass eigenstates to have non-zero masses. 

From studying the end-point of the electron energy distribution in $\beta$-decays, the Mainz and Troitsk experiments have limited the electron neutrino to have a mass smaller than $2.3\eV$ and $2.05\eV$ respectively \citep[95 \% confidence level][]{Kraus:2005,Aseev:2011}. For the specific case of neutrinos being Majorana particles (i.e. their own anti-particles), the Heidelberg-Moscow experiment has limited the effective neutrino mass to be less than $0.35\eV$ (90\% confidence level) using neutrino-less double $\beta$-decay \citep{Klapdor:2006} with similar results from EXO \citep{Auger:2012}. However, no present-day experiment has the sensitivity to measure the absolute neutrino mass. The upcoming KArlsruhe TRItium Neutrino (KATRIN) experiment is expected to improve the sensitivity to $0.2\eV$ (90\% confidence level) from $\beta$-spectroscopy, but this upper limit remains far above the observed mass differences.

The sign of $\Delta m_{32}$ is unknown, which means that we cannot know whether the ordering of the mass states is $m_1 < m_2 << m_3$, similar to the leptons, where $m_e < m_\mu << m_\tau$. This is called normal hierarchy. If $m_3 << m_1 < m_2$, the scenario is called inverted hierarchy.

For the short baseline (SBL) oscillations there are some tensions between the various experiments allowing for, or even preferring, the existence of additional neutrino species that do not interact through the weak force, and are consequently denoted sterile. Depending on the exact analysis, the preferred scenario includes either one or two sterile neutrinos in addition to the three normal ones (3+1 and 3+2), with sterile neutrino masses on the order of $1\eV$ \citep{Kopp:2011,Mention:2011,Huber:2011,Giunti:2011}.

\subsection{Cosmological neutrinos}
\begin{figure}
\centering
	\includegraphics[width=0.7\columnwidth]{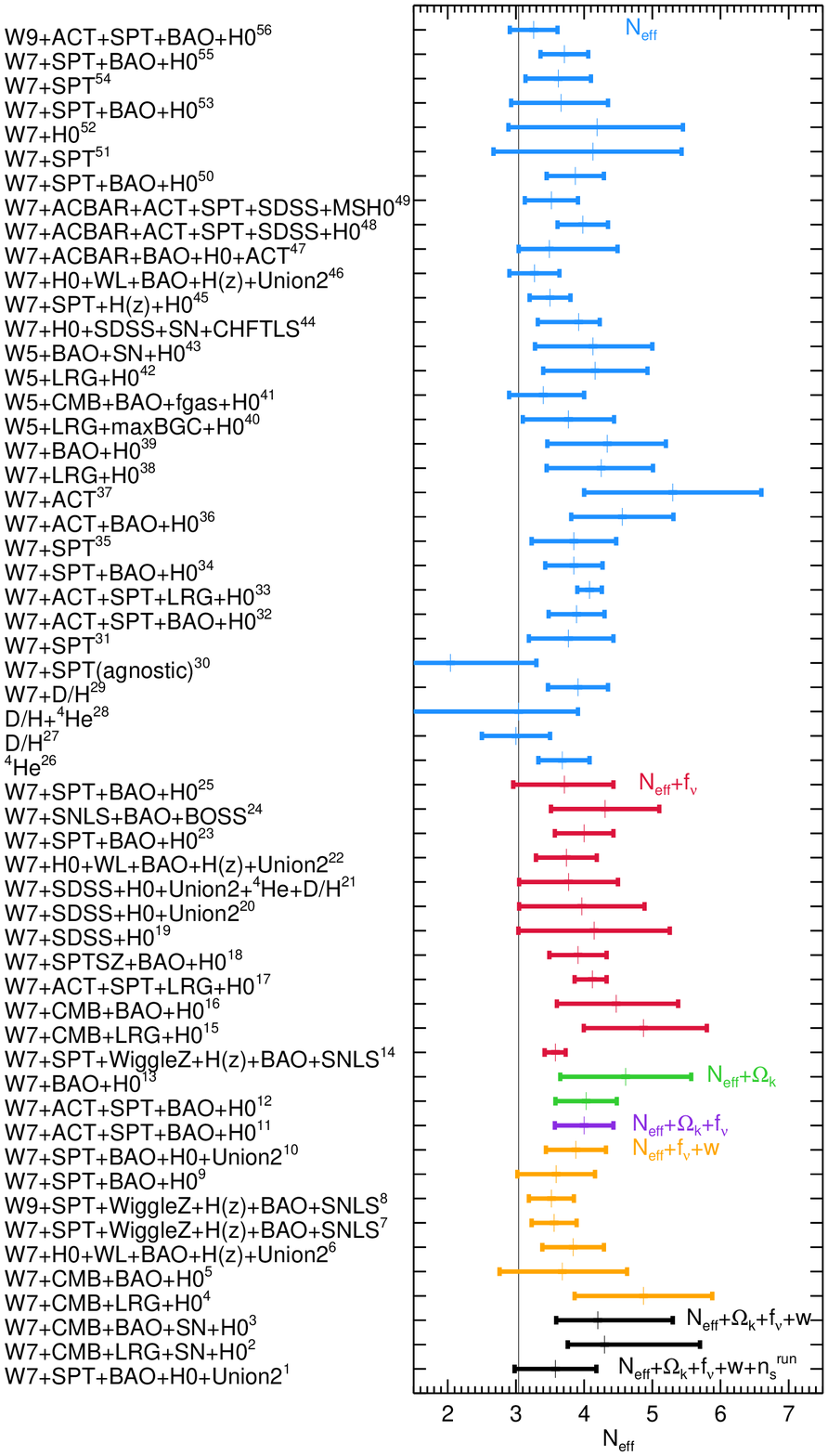} 
	\caption{A selection of cosmological \Neff{} measurements and $68\%$ confidence intervals from the literature for various combinations of models and data sets. W denotes WMAP followed by data release. The models are all \LCDM{} plus the extensions given on the plot. Results from: 
	$^{1, 9, 10, 23, 50}$\citet{Joudaki:2012}, $^{2, 3}$\citet{Gonzalez-Garcia:2010}, $^{4, 5, 13, 15, 16}$\citet{Hamann:2010}, $^{6, 22, 46}$\citet{Wang:2012}, $^{7, 8}$\citet{Riemer-Sorensen:private}, $^{11, 12, 32}$\citet{Smith:2012}, $^{14}$\citet{Riemer-Sorensen:2012b}, $^{17, 33}$\citet{Archidiacono:2011}, $^{18}$\citet{Benson:2012}, $^{19, 20, 21}$\citet{Giusarma:2011}, $^{24}$\citet{Zhao:2012}, $^{25, 51, 52, 53}$\citet{Giusarma:2012}, $^{26}$\citet{Izotov:2010}, $^{27}$\citet{Pettini:2012}, $^{28}$\citet{Mangano:2011}, $^{29}$\citet{Nollett:2011}, $^{30, 31}$\citet{Audren:2012}, $^{34, 35}$\citet{Keisler:2011}, $^{36, 37}$\citet{Dunkley:2011}, $^{38, 39}$\citet{Komatsu:2011}, $^{40, 42, 43}$\citet{Reid:2010}, $^{41}$\citet{Mantz:2010}, $^{44}$\citet{Xia:2012}, $^{45}$\citet{Moresco:2012}, $^{47}$\citet{Gonzalez-Morales:2011}, $^{48, 49}$\citet{Calabrese:2012}, $^{54, 55}$\citet{Hou:2012}, $^{56}$\citet{Hinshaw:2012}.
}
	\label{fig:measurements}
\end{figure}

Observations of the Cosmic Microwave Background (CMB) have, since the 5-year data release from the Wilkinson Microwave Anisotropy Probe (WMAP), displayed a mild preference for an excess of relativistic energy density at the time of decoupling on top of what is provided by the photon density and three Standard Model neutrinos \citep{Komatsu:2009}. This extra radiation has been dubbed `dark radiation' and is often parametrised in terms of extra neutrino species such that $N_\mathrm{eff} = N_\nu + \Delta N$. The most striking result is from the Atacama Cosmology Telescope (ACT) of $N_\mathrm{eff} =  5.3 \pm 1.3$ \citep{Dunkley:2011} which is shown in \figref{measurements} together with a selection of cosmological analyses from the literature. There is a clear trend favouring \Neff{}$>3$ across various models and data sets.

\Neff{} is often referred to as the effective number of neutrino species or the number of relativistic species, but basically it is a parametrisation of anything that would change the expansion rate at the time of decoupling. {Consequently fractional values of \Neff{} do not actually require a non-integer number of neutrino species. Possibilities include other} exotic particles such as axions \citep[e.g.][]{Melchiorri:2007,Nakayama:2011,Erken:2012}, decaying particles or fields \citep[][]{Ichikawa:2007,Nakayama:2011,Fischler:2011,Boehm:2012}, gravity waves \citep{Smith:2006}, extra dimensions \citep{Flambaum:2006}, and early dark energy \citep[and references therein]{Calabrese:2011,Gagnon:2011}.

It is tempting to connect the cosmological preference for \Neff{}$>3$ with the possible additional species from oscillation experiments. In this review we focus on the physical effects (\secref{CosmoEvents}), the cosmological observations (\secref{observables}), the possible origin of these effects in terms of particle physics and other explanations (\secref{discussion}), and what to expect in the future (\secref{future}).

Throughout the paper we use a fiducial cosmology with the best fit WMAP7 parameters from \citet{Komatsu:2011} for illustrations: $\Omega_\mathrm{b,0} = 0.0451$ (baryon density today), $\Omega_{c,0} = 0.226$ (matter density today), $\Omega_{\Lambda,0} = 0.729$ (dark energy density), $H_0 = 70.3 \, \km \, \s^{-1} \, \Mpc ^{-1}$ (Hubble parameter today), $n_\mathrm{s} = 0.966$ (index describing the tilt of power spectrum of primordial density fluctuations), $\Delta^2_\mathcal{R}(k=0.002 \, \Mpc^{-1}) = 2.42 \times 10^{-9}$ (amplitude of the primordial power spectrum at $k=0.002 \Mpc^{-1}$), and unless otherwise stated $w=-1$ (equation of state of the dark energy). All quoted uncertainties on \Neff{} are 68\% confidence levels, and all upper limits on \mnu{} are 95\%.

\section{Early Universe effects} \label{sec:CosmoEvents}
In order to understand the observable effects of dark radiation parametrised as \Neff{}, it is illustrative to follow the normal neutrinos during the evolution of the early Universe depicted for the fiducial cosmology in the left column of the timeline in \figref{timeline}:

\subsubsection*{Neutrino decoupling \\($T_\mathrm{\nu dec} \approx 1.1 \MeV$, $z_\mathrm{\nu dec} \approx 4.5\times10^9$)} 
Neutrinos only interact through the weak interaction and consequently when the early Universe temperature drops below $1 \MeV$, the interaction rate per neutrino drops below the expansion rate, and the neutrinos decouple from the thermal plasma. For neutrino masses below $1\eV$ the neutrinos are still ultra-relativistic when they decouple and they retain a relativistic Fermi-Dirac phase-space distribution. Their energy distribution is described by a blackbody spectrum with a single temperature, and they may have a non-zero chemical potential (related to lepton asymmetry).

\subsubsection*{Big Bang Nucleosynthesis \\ ($T_\mathrm{nuc} \approx0.8\MeV$, $z_\mathrm{nuc} \approx 2.6\times10^{9}$)} 
The formation of deuterium (D) and helium (He) nuclei is affected by the neutrinos in two ways. Firstly, electron neutrinos participate in the charged current weak interactions that determine the neutron to proton ratio (e.g.\ $\nu_e + n \leftrightarrow p + e^-$), and secondly, they increase the expansion rate as
\begin{equation}
H^2(z) \simeq \frac{8\pi G}{3} (\rho_\gamma + \rho_\nu) \, ,
\end{equation}
where $H(z)$ is the Hubble rate and $\rho_\gamma$ and $\rho_\nu$ are the photon and neutrino densities, respectively.
The photon energy density is very well determined from the CMB temperature, so constraints on the expansion in the early Universe, $H(z)$, can be translated directly to $\rho_\nu$ and hence \Neff{}.

\subsubsection*{Photon re-heating (electron recombination) \\ ($T_\mathrm{rec} \approx 0.2 \MeV$, $z_\mathrm{rec} \approx 8.5\times10^8$)}
When the electrons and positrons decouple from the thermal plasma, most of them annihilate. The released energy re-heats the photon population, but not the bulk of the neutrinos since they are no longer in thermal equilibrium with the photons. After annihilation 
\begin{equation} \label{eqn:temp}
T_\nu = (4/11)^{1/3}T_\gamma \, .
\end{equation}
Because the decoupling of the neutrinos is not instantaneous, some of the high energy neutrinos are still in thermal equilibrium with the photons and will feel the re-heating leading to a slight excess of high energy neutrinos relative to a blackbody spectrum. The extra energy per particle is well approximated by increasing the number of particles slightly, and is commonly absorbed into the definition of \Neff{} via
\begin{equation}
\rho_\nu = N_\mathrm{eff}\frac{7\pi^2}{120}T_\nu^4 \, ,
\end{equation}
where $\rho_\nu$ refers to the neutrino energy density when they are still relativistic, and the factor $(7\pi^2/120)T^4_\nu$ is the energy density in one species of relativistic neutrinos. The additional re-heating leads to $N_\mathrm{eff}=3.046$ for $N_\nu = 3$. 

\subsubsection*{Matter-radiation equality \\ ($T_\mathrm{eq} \approx 0.8\eV$, $z_\mathrm{eq} \approx 3200$)}
Relativistic neutrinos behave like radiation, and consequently change the time of radiation-matter equality 
\\\citep{Lesgourgues:2006}
\begin{equation}
\frac{a_\mathrm{eq}}{a_0} = \frac{1}{1-f_\nu}\frac{\Omega_\mathrm{r}}{\Omega_\mathrm{m}}\, ,
\end{equation}
where $\Omega_\mathrm{m}$ is the total matter density, $\Omega_\mathrm{r}$ is the radiation density, and $f_\nu = \Omega_\nu/\Omega_\mathrm{m}$ is the neutrino fraction. The matter-radiation equality determines when the density perturbations can begin to grow and form structures through gravitational collapse. If this is pushed to later times, the peak of the matter power spectrum moves to larger scales (because the particle horizon is larger when the structures begin to collapse).

\subsubsection*{Recombination makes the Universe neutral \\ ($T_\mathrm{rec} \approx 0.3\eV$, $z_\mathrm{eq} \approx 1370$)}
Once the temperature has dropped sufficiently below the ionisation energy of hydrogen, the electrons will couple to the nuclei and the Universe becomes neutral ($z_\mathrm{rec}$ is here defined as when half the electrons are bound). Since the neutrinos have already decoupled a long time ago, they are unaffected by this, but the recombination is the beginning of the CMB era.

\subsubsection*{Photon decoupling and last scattering \\ ($T_\mathrm{*} \approx T_\mathrm{ls} \approx 0.25 \eV$,  $z_\mathrm{*} \approx z_\mathrm{ls} \approx1100$)}
When the temperature of the Universe has dropped to $\approx 0.25 \eV$, the collision rate between photons and free electrons drops below the expansion rate and photons propagate freely. The last scattering between photons and electrons is observed today as the CMB. The neutrinos do not care about this, but the CMB is one of the most powerful probes available to study the early Universe (see \secref{CMB}).

\subsubsection*{Baryon decoupling and the drag epoch \\ ($T_\mathrm{drag} \approx 0.24-0.25 \eV$, $z_\mathrm{drag} \approx1020-1060$)}
Even after the photon-electron scattering ceases and the mean free path of photons becomes effectively infinite, a small number of the photons do still interact with the far less numerous baryons ($n_\mathrm{b}/ n_\gamma \approx 10^{-10}$), prolonging the pressure wave in the baryons. This is essentially a difference between when the photons stop noticing the baryons (decoupling), and when the baryons stop noticing the photons. This delay is known as the drag epoch, at the end of which the density fluctuations that form the seeds of structure are frozen in. This event does not affect the neutrinos, but the decoupling allows the baryons to cluster in the dark matter potentials initiating the formation of observable structures.

\subsubsection*{Becoming non-relativistic \\ ($T_\mathrm{nr}\approx 0.05\eV$, $z_\mathrm{nr}\approx200$ depending on \mnu)}
At least two of the neutrino mass eigenstates are massive so their velocities will decay adiabatically with the expansion of the Universe, and eventually they become non-relativistic ($m_\nu c^2> p_\nu c$) \citep{Lesgourgues:2006, deBernardis:2009}
\begin{equation}
z_{\rm nr} \approx 2\times10^3 \frac{m_{\nu,i}}{\eV} \, . 
\end{equation}
The last eigenstate is allowed to be massless and if this is the case, it will remain relativistic forever.
When the neutrinos become non-relativistic they behave like a species of warm/hot dark matter, slowing the growth of density fluctuations on scales smaller than their free-streaming length\footnote{The free-streaming length is the average distance the neutrinos propagate freely, and for kinematic reasons neutrinos cannot be confined (or kept out of) regions smaller than the free-streaming length.} (see \secref{LSS}).

\begin{figure}
\centering
	\includegraphics[width=0.55\columnwidth]{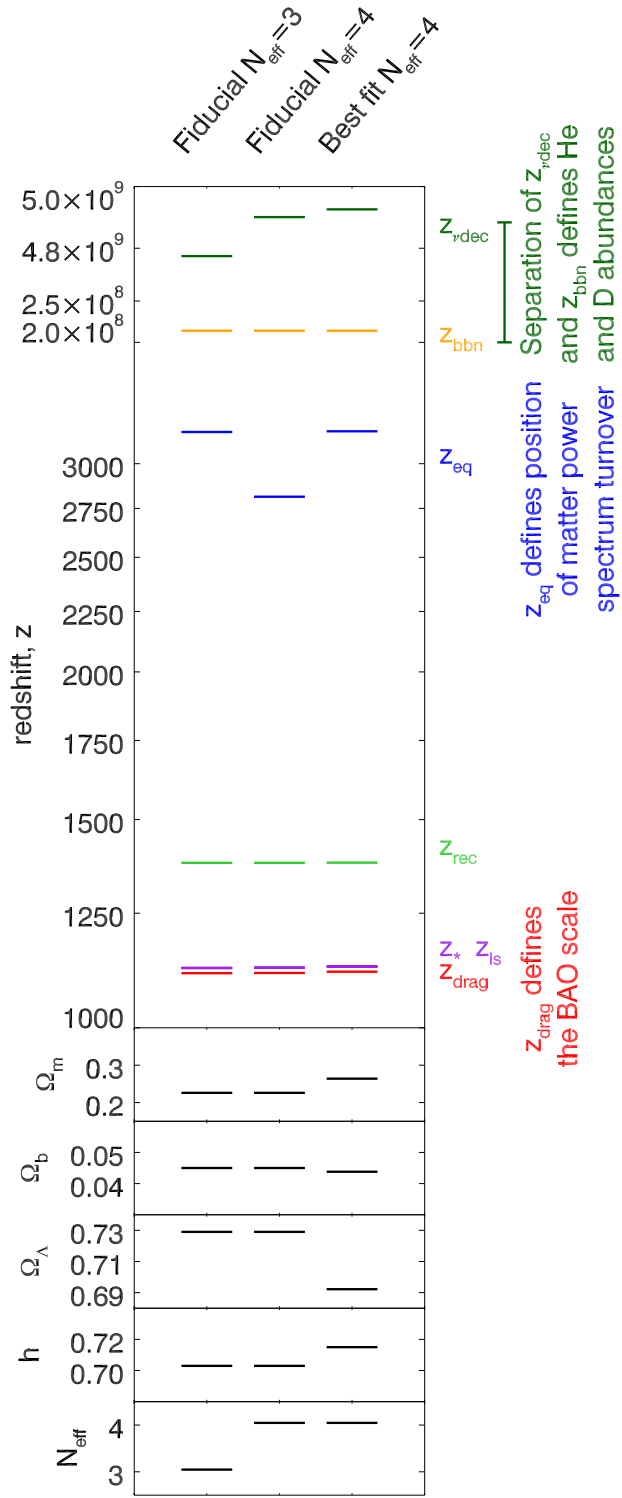} 
	\caption{The redshifts of important events in the early Universe (described in the main text) and their dependence on \Neff{} (the first two columns) and on cosmological parameters for fixed \Neff=4 (the second and third columns). Changing \Neff{} mainly changes the redshift of matter-radiation equality ($z_\mathrm{eq}$) for fixed cosmological parameters. However, this change can be compensated by increasing the values of \hh{} and \omm{}. Notice that the neutrino decoupling remains early allowing one to distinguish between the first and third scenarios through deuterium and helium abundances.}
	\label{fig:timeline}
\end{figure}

\section{Observable effects} \label{sec:observables}
\begin{figure*}
 \centering
	\includegraphics[width=0.95\textwidth]{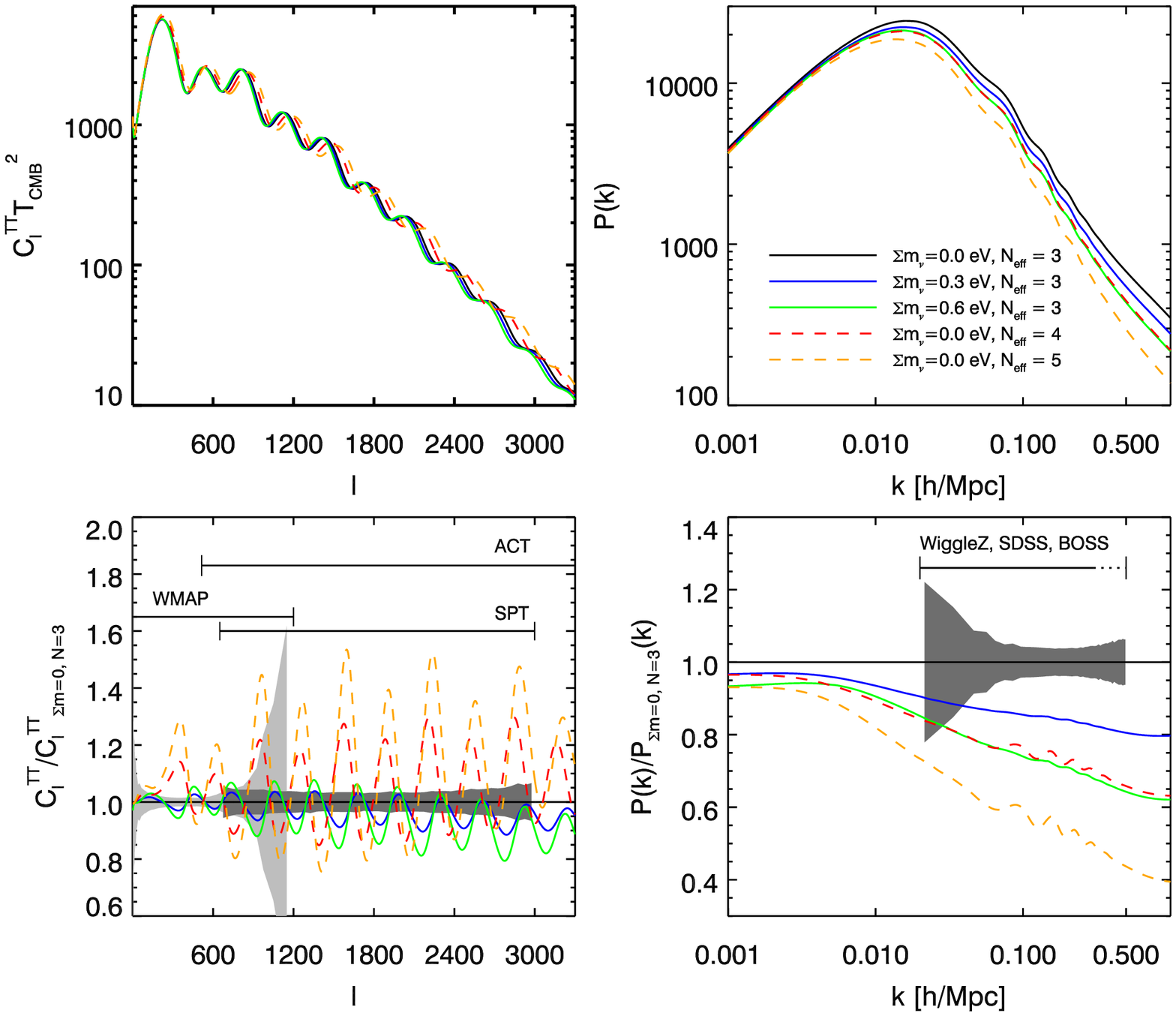} 
	\caption{Illustration of the CMB (left) and matter (right) power spectra for a fiducial cosmology and how they change for varying neutrino mass (solid lines, we vary $f_\nu = \Omega_\nu/\Omega_m = \sum m_\nu/(93.14 \eV \Omega_\mathrm{m} h^2)$), and effective number of neutrinos (dashed lines) fixing all other parameters. The bottom two plots show the ratio of the power spectra when varying \mnu{} and \Neff{} relative to \mnu=$0\eV$, \Neff=0.
			 \mnu{} does not affect the CMB power spectrum much, but changes the matter power spectrum. The effect of \Neff{} is clearly visible for small scales (high values of $l$) in the CMB power spectrum, and the two parameters are clearly degenerate for the matter power spectrum unless the peak position is very precisely measured. The horizontal lines indicate the coverage by current and near future experiments, and the shaded areas indicates the relative magnitude of the uncertainties for WMAP (light grey) and SPT (dark grey) in the CMB plot, and WiggleZ (dark grey) for galaxy surveys.}
	\label{fig:powerspec}
\end{figure*}

The main effect of dark radiation is to alter the expansion rate and the time of matter-radiation equality, which affects nucleosynthesis, the CMB, and structure formation. The relevant observables for nucleosynthesis are the deuterium and helium abundances, for the CMB it is the power spectrum of temperature and polarisation fluctuations expressed in terms of spherical harmonics, and for structure formation it is the (matter) density power spectrum inferred from e.g.\ galaxy surveys. \figref{powerspec} illustrates the effects of increasing the neutrino mass (solid lines) and number of species (dashed lines) on the CMB and density power spectra. In the following sections we will discuss these observable effects in detail.

What is probed directly by cosmological analyses is not the neutrino mass, but the neutrino density, $\rho_\nu$, which can be expressed in terms of the sum of neutrino masses \citep{Lesgourgues:2006}
\begin{equation}
\Omega_\nu = \frac{\rho_\nu}{\rho_c} = \frac{\sum_{i=1}^{N_\nu} m_{\nu,i}}{93.14\eV h^2} \, ,
\end{equation}
where $\rho_c$ is the critical energy density for a flat Universe. Because of the small-ness of the measured mass differences relative to the upper limits, it is reasonable to assume that if the individual neutrino masses are near the upper limit, they are effectively equal. If the individual masses are close to the lower limit, the hierarchy will play an important role \citep{Jimenez:2010}.

\subsection{Big Bang Nucleosynthesis}
The presence of dark radiation mainly alters the nucleosynthesis by increasing the expansion rate leading to higher deuterium and helium abundances than are observed today \citep{Lesgourgues:2006,Steigman:2012}.

\subsubsection{Helium}
The freeze-out of weak interactions determines the neutron/proton ratio, that will eventually determine the helium-fraction ($Y_p = \rho_\mathrm{He}/\rho_\mathrm{baryon}$), since all available neutrons either become bound in helium or decay. Increasing the expansion rate will induce earlier freeze-out of the weak interaction (see \figref{timeline}), and more neutrons will be available leading to a higher helium fraction.

Helium-4 (\He{}) is the nucleosynthesis relic most sensitive to \Neff{} but because it is produced in stars it is not trivial to obtain a robust value of the primordial abundance. High quoted values of \Neff{} from nucleosynthesis are usually driven by the adopted value of $Y_p$ and its uncertainty. Since the beginning of the 1990s the observed value has increased from $Y_p=0.2462\pm 0.0015$ \citep{Izotov:1999} to $Y_p=0.2565\pm 0.001 \mathrm{(stat.)}\pm0.005\mathrm{(syst.)}$ \citep{Izotov:2010} which drives the value of \Neff{} up. 

\subsubsection{Deuterium}
The deuterium abundance is determined by how much deuterium is fused into \He{} after deuterium formation has ceased because of too low neutron and proton densities towards the end of the nucleosynthesis. Increasing \Neff{} willl speed up the expansion (but not as much as when helium formed) allowing for less deuterium to be fused into helium and consequently leading to a higher deuterium abundance \citep{Nollett:2011}. 

\subsubsection{Lithium}
The lithium abundance predicted from standard nucleosynthesis is a factor of three higher than that observed in metal poor halo stars. Increasing \Neff{} does not change this problem \citep{Steigman:2012}.

\subsubsection{Results from Nucleosynthesis}
There have been a number of recent attempts to derive \Neff{} from nucleosynthesis alone e.g.\ \citet{Mangano:2011,Pettini:2012}, which all seem to be consistent with $N_\mathrm{eff} = 3$. Analyses combining nucleosynthesis and CMB appear to prefer $N>3$ \citep[e.g.][]{Nollett:2011,Hamann:2011}.

\subsection{Cosmic Microwave Background} \label{sec:CMB}

The distribution of hot and cold spots in the CMB is given by the fluctuations in the photon-baryon plasma at the surface of last-scattering. These fluctuations are generated by sound waves in the plasma, where the repulsive electrostatic force is balanced by the gravitational attraction of dark matter potentials that have been growing since $z_{\rm eq}$ (see \figref{timeline}). They manifest in the observed CMB power spectrum as actual oscillations, as shown in \figref{powerspec}. Accurate measurement of the positions and heights of the peaks and troughs can give us two important pieces of information about the relativistic energy density.

Firstly, the amplitude of the oscillations relates to the matter-to-radiation density ratio. Pressure support of the compression wave in the plasma causes the gravitational potential to decay, enhancing the amplitude of these perturbations. This enhancement occurs only for perturbations that have already crossed the horizon before matter-radiation equality \citep{HuSugiyama96}, which is true for the third and subsequent peaks. This is why the third peak is enhanced with respect to the second, and the amplitude ratio of the first to third peaks gives a good measurement of $z_{\rm eq}$, and so \Neff{}. 

Secondly, the peak positions give the size of the sound horizon at decoupling, namely the maximum distance the acoustic oscillations can travel before the sound speed in the fluid drops to zero. This relates directly to the Hubble rate, and so is also influenced by \Neff{}. Notice from \figref{powerspec} that increasing the value of \Neff{} moves the acoustic peaks to smaller scales (since the sound horizon, $r_s$, becomes smaller for increasing \Neff), whereas in the matter power spectrum the turnover moves to larger scales (related to the change of $z_{\rm eq}$ seen in \figref{timeline}). This anti-correlation between effects in the acoustic peaks in the CMB and the turnover in the matter power spectrum provides an important consistency check between the early (pre-CMB) and late time (post-CMB) physics.

In addition, we must consider the damping of the CMB power spectrum `tail' on small scales. Photons in the photon-baryon fluid will diffuse out of over-dense regions, and generate an anisotropic stress that damps the oscillations on small scales ($\ell > 200$). This effect is called Silk or diffusion damping. The neutrinos contribute to the small-scale damping both indirectly -- by changing the expansion and thereby the damping from the photons -- and directly -- through their own free-streaming. 

Increasing the expansion rate by adding extra relativistic material will increase the Silk damping because structures have less time to grow and are consequently easier to wipe out through diffusion. Similarly, increasing the expansion rate increases the helium abundance, which also increases the damping because the helium abundance controls the number of free electrons per unit baryon mass.

Neutrinos contribute to the damping through anisotropic stress and free-streaming effects \citep{HSSW96}. Similar to the photons, the neutrinos have a viscosity that creates an anisotropic stress and consequently generates damping on small scales. Fluctuations in the neutrino density lead to a shift in the positions of the acoustic peaks. These effects are non-negligible, as at the time of decoupling, the standard neutrinos contribute around 10\% of the total energy density of the Universe.
Because the neutrinos are free-streaming, they suppress structure formation on scales smaller than their free-streaming length (see \secref{LSS}). The combination of these effects tell us something more than simply the density of the relativistic species, and can provide details on both the mass and viscosity of the dark radiation.

\subsubsection{Results from CMB}
CMB observations of the first three acoustic peaks in the temperature power spectra from WMAP have accurately measured the redshift of matter-radiation equality to be $z_{\rm eq} = 3145^{+140}_{-139}$ \citep{Komatsu:2009}, for models where \Neff{} is allowed to vary. However, WMAP cannot measure the tail of the CMB temperature power spectrum ($\ell > 1200$), and so is unable to put an upper limit on \Neff{}, giving only a lower limit of $N_{\rm eff} > 2.7$ (95\% confidence level). An upper limit can be found either by combining WMAP with late-time measurements of the matter density (SN, BAO, or galaxy power spectra), or through measurements of the CMB tail. When combined with small-scale CMB experiments such as the Atacama Cosmology Telescope (ATC) or South Pole Telescope (SPT), the constraints improve (see \figref{measurements}), giving $N_{\rm eff} = 5.3 \pm 1.3$ from WMAP + ACT \citep{Dunkley:2011} and $N_{\rm eff} =3.85 \pm 0.62$ from WMAP + SPT \citep{Keisler:2011}.

In terms of the neutrino mass, the CMB alone cannot provide very strong constraints. In order to have a measurable effect on the CMB, the neutrinos would have to be non-relativistic before photon decoupling, but current cosmological limits indicate that they were relativistic at this time. The upper limit imposed by WMAP is $\sum m_{\nu} < 1.3 \eV$ \citep{Komatsu:2009}, which is close to (in fact beyond!) the absolute limit of $\sum m_{\nu} < 1.5 \eV$ one can theoretically expect to derive from CMB alone \citep{Ichikawa05}. However, the CMB does play an important role of providing good `base-level' constraints on the matter density ($\Omega_m$) and the amplitude of primordial fluctuations ($\Delta^2_\mathcal{R}$), which need to be jointly measured when combining the CMB with other cosmological datasets to measure the neutrino mass.

\subsection{Large Scale Structure} \label{sec:LSS}
Like the CMB, the large scale structure is affected by neutrinos through the expansion and their free-streaming effects. The expansion determines the time available for sound waves to propagate in the baryon-photon fluid before recombination, which will imprint on the BAO scale measurable from both the CMB and in the galaxy power spectrum (see \figref{timeline}). Increasing \Neff{} pushes the BAO peak towards smaller scales, indicated by a slight shift of the positions of the `wiggles' in the galaxy power spectrum in \figref{powerspec}.  The position of the matter power spectrum turnover is determined by the time available for structures to form, and is consequently pushed to smaller scales as matter-radiation equality moves to later times with increasing \Neff. However, this effect is degenerate with effects from the ordering of the neutrino masses (the hierarchy) that dictates whether one third, or two thirds of the neutrinos will be `heavy' or `light'. The light ones will be radiation-like for longer and enhance structure growth on large scales \citep{Jimenez:2010,Wagner:2012}.

Massive neutrinos (or other particles) will also slow down structure formation due to their free-streaming properties. After decoupling, their masses will lead to an adiabatic decay of their thermal velocities as \citep{Komatsu:2011}
\begin{equation}
v_\mathrm{thermal}=151(1+z)(1\eV/m_\nu)\km\s^{-1} \, .
\end{equation}
Eventually they become non-relativistic and behave as a species of warm/hot dark matter, suppressing growth of density fluctuations on scales characterised by their free-streaming length at the time when they become non-relativistic \citep{Lesgourgues:2006}
\begin{eqnarray}
k_\mathrm{FS} 	&=& \sqrt{\frac{3}{2}}\frac{H(t)}{v_\mathrm{thermal}(t)(1+z)} \\
			&\approx& 0.82\frac{\sqrt{\Omega_\Lambda+\Omega_m(1+z)^3}}{(1+z)^2}\left(\frac{m_\nu}{1\eV}\right)h \Mpc^{-1} \, . \nonumber
\end{eqnarray}
This happens because the neutrinos cannot cluster on scales smaller than their free-streaming length and consequently the density fluctuations are suppressed by a constant factor proportional to the neutrino density, and will grow more slowly than without the presence of massive neutrinos. The effect is observable in the density power spectrum as a suppression of structure on small scales shown in the right part of \figref{powerspec}. However, the presence of neutrinos also enhances the density perturbations on large scales (by adding to the radiation density for fixed \omm{}), which boosts structure formation in general \citep{Hou:2011}.

The evolution of the density perturbations can be predicted by solving the coupled set of Einstein and Boltzman equations. For small density perturbations the linearised equations are sufficient (calculated numerically e.g.\ using CAMB\footnote{\url{http://camb.info}} or CLASS\footnote{\url{http://lesgourg.web.cern.ch/lesgourg/class.php}}), but when the density perturbations grow sufficiently large, the higher-order, non-linear terms become important and have to be included, which is computationally very intensive (and not yet solved analytically). Instead one has to rely on second order perturbation theory \citep[e.g.][]{Saito:2009,Taruya:2012} or simulations \citep[e.g.][]{Smith:2003,Jennings:2010}. Numerical solutions show that for $f_\nu=\Omega_\nu/\Omega_m<0.07$ the suppression is $\delta P/P =-8f_\nu$ for linear structure formation \citep{Hu:1998} and the effect increases for non-linear structure formation \citep[$8\rightarrow9.6$][]{Brandbyge:2008, Brandbyge:2009, Brandbyge:2010, Viel:2010, Agarwal:2011}.

The neutrinos affect the total matter distribution which is dominated by the dark matter, and consequently not directly observable. Instead we use tracers such as galaxies or the absorption lines from neutral hydrogen clouds along the line of sight to distant quasars (Lyman alpha forest), for which the power spectrum may differ slightly from the total matter power spectrum if baryons do not perfectly trace the dark matter. Depending on the tracer, the bias, $b$ defined as $P_\mathrm{obs}=b^2P_\mathrm{m}$, might be scale dependent or scale-independent \citep{Schulz:2006}. A scale dependent bias can introduce a degeneracy with the slope change induced by the neutrinos. 

For a more in-depth, technical account of massive neutrinos in cosmology, see \citet{Lesgourgues:2006} or \citet{Wong:2011}. 

\subsubsection{Results from large scale structure}
Large scale structure data alone cannot constrain all the cosmological parameters, so they are usually analysed jointly with CMB data. Currently there is no precise measurement of the matter power spectrum turnover \citep{Poole:2012b}, so large scale structure data mainly improves on the \mnu{} constraints, and are consequently most important when fitting for \mnu{} and \Neff{} simultaneously rather than for \Neff{} alone. The current limits of \mnu$<0.3\eV$ for \LCDM{} + \mnu{} \citep{Moresco:2012,Xia:2012,dePutter:2012,Riemer-Sorensen:2012} and \mnu{}$<0.65\eV$ for \LCDM{} + \mnu{} + \Neff{} \citep[e.g.][]{Hamann:2010,Joudaki:2012,Riemer-Sorensen:2012b,Wang:2012} are stronger than what is provided by particle physics experiments.

\section{Discussion} \label{sec:discussion}
\subsection{Do we need \Neff{}?}
Perfect measurements of the standard \LCDM{} parameters do not exist. Adding \Neff{} to the cosmological model does not shift the best fit values of the standard \LCDM{} parameters in any systematic way, indicating that the preference for high \Neff{} is not due to an off-set measurement of a single parameter, but can be accommodated within the uncertainties of the standard parameter values.

\begin{figure*}
\centering
	\includegraphics[width=0.95\textwidth]{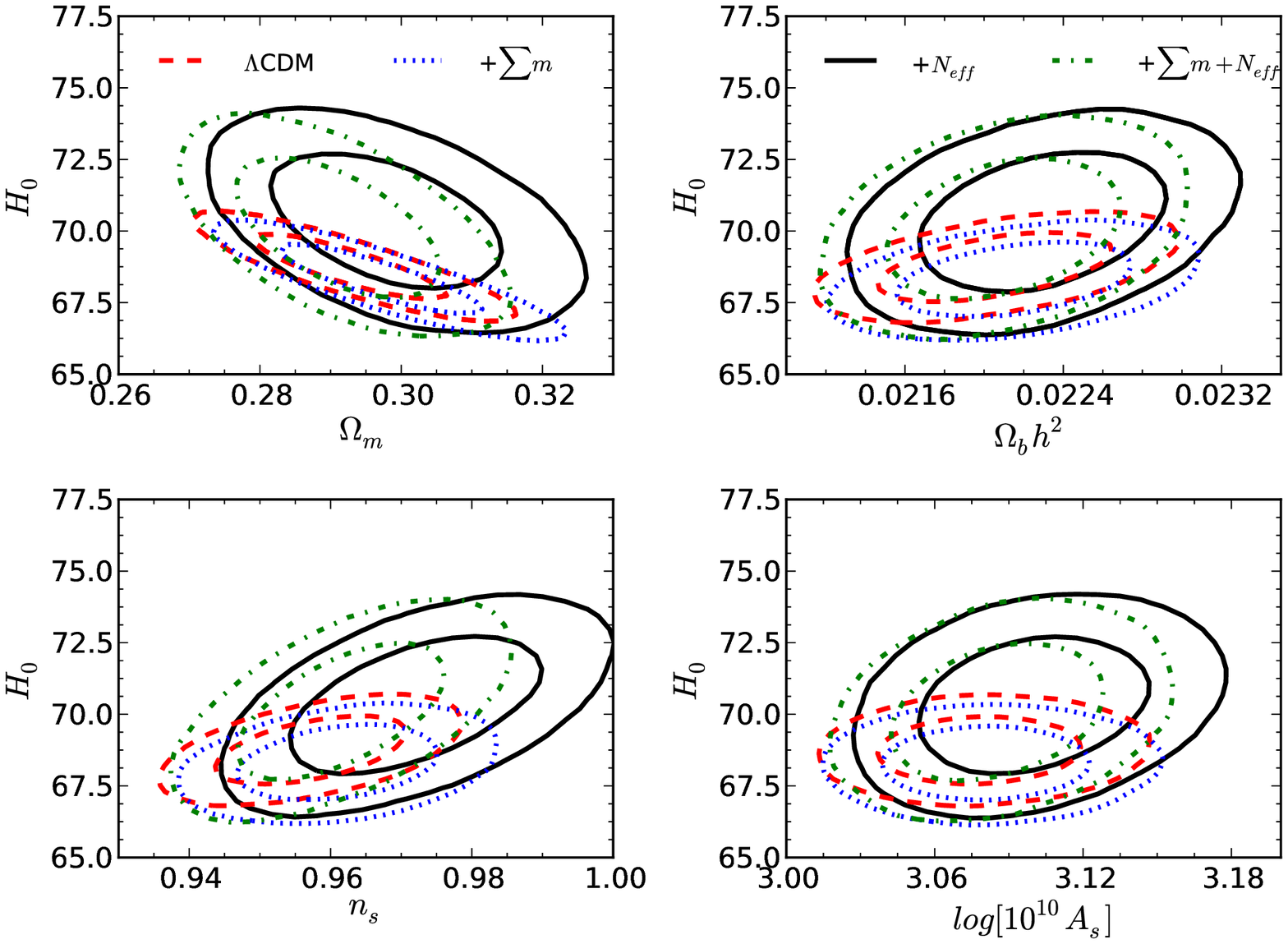} 
	\caption{The 68\% and 95\% confidence limit contours for fitting CMB (WMAP7+SPT) + large scale structure (WiggleZ) + $H(z)$ + BAO (SDSS, 6dFGS, BOSS) + SN (SNLS) data with \LCDM{} (dashed red) and extensions to the model with \mnu{} (dashed blue), \Neff{} (dot-dashed green), and both \mnu{} and \Neff{} (solid black) from \citep{Riemer-Sorensen:2012b}. The resulting contours are consistent with each other for all parameters clearly showing that the preference for high values of \Neff{} is not related to an off-set in the measurement of a single standard \LCDM{} parameter, but can be added within the standard parameter uncertainties.}
	\label{fig:LCDM}
\end{figure*}

\figref{LCDM} compares the contours from fitting a range of cosmological data (see caption) with a standard \LCDM{} model alone, and with \mnu{} and \Neff{} extensions individually and combined. Naturally the constraints are tighter for pure \LCDM, since there are fewer parameters to constrain, but overall the constraints are consistent for all parameters: Adding \mnu{} to the model (dotted blue) does not change the \LCDM{} parameters or uncertainties significantly whereas adding \Neff{} (dot-dashed green) increases the values and uncertainties, particularly for $H_0$. Adding both parameters simultaneously (solid black) increases the preferred values of $H_0$ and $n_s$ but also the uncertainties, so the values remain consistent with the pure \LCDM{} case. The shift in $n_\mathrm{s}$ happens because the increased expansion ($H_0$) changes the height of the peaks in the CMB, as it corresponds to increasing the physical matter density. Increasing \Neff{} recovers the details of the CMB peaks (because the original ratio of matter to radiation is recovered, restoring the details of the Silk damping), but with too much power on large scales, which can be eliminated shifting the primordial power spectrum from very red ($n_s=0.96$) to slightly less red ($n_s=0.98-1.0$). 

\subsection{Priors}
Since most constraints on \Neff{} are obtained through Bayesian analyses, one could be concerned that the preference for \Neff$>3$ originates in the choice of priors on the model parameters. Both \citet{Gonzalez-Morales:2011} and \citet{Hamann:2012} investigated this and found no significant shifts in the \Neff{} constraints for various selections of priors.

\subsection{Beyond \LCDM+\mnu+\Neff{}}
Other physical effects such as curvature, varying equation of state, running of the spectral index etc. could mimic \Neff{}$>3$ if not properly accounted for in the modelling. \citet{Joudaki:2012} demonstrated that the deviation from $N_\mathrm{eff}=3.046$ is diminished if allowing for curvature, varying equation of state, running of the spectral index, and/or the helium fraction by fitting to a combination of CMB + BAO + {\it HST} + SN data. However, for all parameter combinations the preferred value of \Neff{} was still above three (see \figref{measurements}), and only when more than one extra parameter extension of the $\Lambda$CDM cosmology (e.g.\ curvature {\it and} varying equation of state in addition to \Neff{} and \mnu{}) were considered, the preferred value of \Neff{} became consistent with three within one standard deviation. Consequently \emph{the preference for $N_\mathrm{eff}\neq3$ does not appear to be strongly dependent on the complexity of the cosmological model}. A similar conclusion was reached by \citet{Menegoni:2012} which concluded that even when varying both the helium fraction and the fine structure constant on top of \LCDM{} + \Neff{} the preferred value of \Neff{} remained high.

\citet{Audren:2012} took an alternative approach and separated cosmological effects originating directly from early Universe conditions from those related to later time effects, when analysing CMB data. Marginalising over the late time effects, they found the preferred value of \Neff{} to be consistent with three, and consequently suggested that the preference for high values of \Neff{} does not relate to changed expansion in the early Universe, but rather to our understanding and modelling of late time effects.

\subsection{Thermalisation} \label{sec:thermalisation}
All current analyses relating the dark radiation to neutrinos rely on an assumption of the sterile neutrinos being in thermal equilibrium with the photons before the neutrinos decouple. However, this is an ad-hoc assumption and a full thermalisation calculation involving momentum-dependent density matrices for three active and $\Delta N$ sterile flavours has never been performed \citep{Abazajian:2012}.

The thermalisation can be suppressed in the presence of large lepton asymmetry \citep[more leptons than anti-leptons,][]{Foot:1995}. In that case the mass constraints of \mnu$<0.3\eV$ from large scale structure do not apply because they rely on the assumption of thermalisation \citep{Hannestad:2012}. This allows for a joint interpretation of the cosmology and particle physics results under very specific early Universe conditions. The necessary lepton asymmetry ($L=(n_{f} - n_{\overline{f}}) /n_\gamma \approx10^{-2}$) is below any observational constraints, but well above the usually assumed value which is of the same order of magnitude as the baryon asymmetry ($L\approx10^{-10}$). Large neutrino asymmetries can be generated by the active-sterile neutrino mixing without the need to invoke new physics \citep{Foot:1996}.

\subsection{Non-standard decoupling}
The number of relativistic degrees of freedom depends on the decoupling processes involved. In the three-neutrino framework, the non-instantaneous decoupling of the standard model neutrinos leads to an expected value of \Neff=3.046 for standard decoupling. However, non-standard decoupling can maximally increase this value to \Neff=3.12 for three neutrino species \citep{Mangano:2006}, and consequently non-standard decoupling cannot alone explain the preference for the currently observed high values of \Neff{}.

\subsection{Particle properties}
The value of \Neff$>3$ is often interpreted as an extra specie of non-interacting relativistic neutrino-like particle. If this is the case, it should behave as a neutrino during the evolution of density perturbations which is can be characterised by a relation between its effective sound speed and viscosity parameter of $c_\mathrm{eff}^2 = c_\mathrm{vis}^2 = 1/3$ \citep{Smith:2012}.  $c_\mathrm{vis}$ describes the ability of the particles to free-stream out of gravitational potentials, and $c_\mathrm{eff}\neq1/3$ allows for the existence non-adiabatic pressure perturbations. If the dark radiation is composed of a particle with significant interaction, these parameters can take other values \citep{Basboll:2009}, thus allowing us to distinguish between hypothesised species \citep{Trotta:2005}.
This was tested by \citet{Archidiacono:2011} using a combination of CMB + large scale structure + {\it HST} data. They found $N_\mathrm{eff}>3$ at $2\sigma$ $(N_\mathrm{eff} = 4.15^{+0.19}_{-0.23})$ and $c_\mathrm{eff}^2=0.24^{+0.03}_{-0.02}$ consistent with $1/3$ while $c_\mathrm{vis}^2$ remains unconstrained. This result is consistent with a relativistic non-interacting particle but is not a unique signature of such a particle.

For the specific case of hypothetical sterile neutrinos there is another effect to consider, which could change the nucleosynthesis, namely the distortion of the electron-neutrino phase space distribution through neutrino oscillations. For the parameters favoured by short baseline experiments this effect is negligible because the sterile and active neutrinos will be in thermal equilibrium and consequently share the same phase space distribution.

\subsection{Time evolution of \Neff}
There have been a number of attempts to derive \Neff{} from nucleosynthesis alone \citep[e.g.][]{Mangano:2011,Pettini:2012} that all seem to be consistent with $N_\mathrm{eff} = 3$, while analyses combining nucleosynthesis and CMB seem to prefer $N>3$ \citep[e.g.][]{Nollett:2011,Hamann:2011}. It is concerning that the preference for \Neff$>3$ is present in all analyses including CMB, but not preferred by nucleosynthesis alone. This could indicate a systematic error in one of the data sets. However, nucleosynthesis alone relies on very few and notoriously difficult measurements of the deuterium and helium fractions. 

The difference between the results could also be interpreted as potential tension between \Neff{} measured at two different epochs (nucleosynthesis and recombination), where we naively would expect \Neff{} to be the same. A temporal variation could be explained theoretically by particles decaying to photons or neutrinos and thereby altering their temperature distributions \citep{Ichikawa:2007,Fischler:2011,Bjaelde:2012,Hooper:2012,Menestrina:2012}, or heavier particles in thermal equilibrium with the neutrinos, but not the photons \citep{Boehm:2012}. Currently there is no experimental evidence for the existence of such particles.

\subsection{Link between particle physics and cosmology} \label{sec:ParticleResults}
Is the cosmological dark radiation related to the sterile neutrinos from the oscillation experiments?

The neutrino oscillation results favour a mass of order $1\eV$, which is incompatible with cosmological mass constraints from combinations of CMB and large scale structure measurements \citep[e.g.\ $\sum m_\nu <0.3-0.6 \eV$, ][]{Riemer-Sorensen:2012}.  \citet{Archidiacono:2012} analysed the short baseline oscillation data together with cosmological data and found that short baseline data and CMB are compatible with both 3+1 and 3+2 scenarios, but when adding large scale structure both scenarios are under serious pressure. However, it has been shown that factors such as initial lepton asymmetry (\secref{thermalisation}) can alleviate these constraints by introducing a non-thermal neutrino spectrum for which the neutrino mass constraints from large scale structure can be avoided \citep{Hannestad:2012}.

\section{Future measurements of \Neff} \label{sec:future}
Due to its high spatial resolution and several spectral bands, the Planck microwave observatory, is expected to measure \Neff{} with a $1\sigma$ uncertainty of $0.20-0.25$ \citep{Bashinsky:2004,Hannestad:2006,Poole:2012b}. The neutrino anisotropy will also lead to a polarisation of the CMB detectable in future experiments. Combining Planck with future small-scale polarisation experiments, a sensitivity of 0.05 may be achievable \citep{Galli:2010,CoRE:2011}, which would almost allow for a measurement of the thermal distortion of the neutrino spectrum (the 0.046).

Measuring the position of the peak of the matter power spectrum (the turnover) would give another handle on \Neff{}. \citet{Poole:2012b} predicts that a Euclid-like galaxy survey will be able to constrain \Neff{} with a precision of 20\% independently of the CMB. However, the position of the turnover can be degenerate with neutrino hierarchy effects \citep{Wagner:2012}. The degeneracies between \mnu{}, \Neff, and the hierarchy allows for a measurement of either of them if the remaining parameters can be `fixed' by independent data.

Any neutrino-like behaving particle (including sterile neutrinos and axions etc., which decouple early as relativistic and become non-relativistic) can mimic the effect of the neutrinos. If the value of $N_\mathrm{eff}>3$ is due to a particle, the result points towards physics beyond the Standard Model, and its existence it will have to be confirmed by particle physics laboratory experiments. If no compatible particles are found in laboratory experiments, the cosmological preference for additional species indicates a lack of understanding and modelling of cosmological data, both of which are interesting scenarios.

\section{Conclusions}
Cosmological data seems to favour the presence of dark radiation at the time of photon decoupling at a $2\sigma$ level, while short baseline particle physics experiments are best explained by the existence of 1-2 sterile neutrinos. Unfortunately the details and required particle properties of the two scenarios are hard to reconcile, and appear to have separate solutions. No matter whether the cosmological trend is due to new particles or systematics, it is a clear indication of something we do not understand.

\section*{Acknowledgments}
We would like to thank Chris Blake and Julien Lesgourgues for constructive discussions and the anonymous referee for useful comments adding quality to the review.
\bibliographystyle{apj}


\end{document}